\documentclass[twocolumn,english,preprintnumbers,amsmath,amssymb,pre]{revtex4}

\usepackage{graphicx}
\usepackage{color}

\begin{document}

\title{Quantum effects in two-dimensional silicon carbide}
\author{Carlos P. Herrero}
\author{Rafael Ram\'irez}
\affiliation{Instituto de Ciencia de Materiales de Madrid (ICMM),
         Consejo Superior de Investigaciones Cient\'ificas (CSIC),
         Campus de Cantoblanco, 28049 Madrid, Spain }
\date{\today}

\begin{abstract}
Two-dimensional (2D) silicon carbide is an emergent direct band-gap
semiconductor, recently synthesized, with potential applications in 
electronic devices and optoelectronics.
Here, we study nuclear quantum effects in this 2D material
by means of path-integral molecular dynamics (PIMD) simulations
in the temperature range from 25 to 1500~K. Interatomic interactions 
are modeled by a tight-binding Hamiltonian fitted to density-functional 
calculations. 
Quantum atomic delocalization combined with anharmonicity of the
vibrational modes cause changes in structural and thermal properties 
of 2D SiC, which we quantify by comparison of PIMD results with 
those derived from classical molecular dynamics simulations,
as well as with those given by a quantum harmonic approximation.
Nuclear quantum effects are found to be appreciable in structural 
properties such as the layer area and interatomic distances. 
Moreover, we consider a {\em real} area for the SiC sheet, which
takes into account bending and rippling at finite temperatures.
Differences between this area and the in-plane area are discussed
in the context of quantum atomic dynamics. 
The bending constant ($\kappa = 1.0$ eV) and the 2D modulus of
hydrostatic compression ($B_{xy}$ = 5.5 eV/\AA$^2$) are clearly 
lower than the corresponding values for graphene.
This study paves the way for a deeper understanding of the 
elastic and mechanical properties of 2D SiC.  \\

\end{abstract}

\maketitle

\section{Introduction}

Bulk silicon carbide has been known for many years as a material
with remarkable physical properties, such as low density, high thermal
conductivity, low thermal expansion, high strength, and high
refractive index \cite{me07c}.
This material is known in more than 250 different polytypes, 
many of them with hexagonal crystalline structure.
In recent years, several new materials containing both silicon and
carbon have been studied. Among them, one finds fullerenes, 
nanotubes, and two-dimensional (2D) structures 
\cite{me07c,hs11,sh15,be10b}.
In particular, great progress has been lately made in the
understanding and synthesis of 2D SiC \cite{li12,ch16,hu19,ch21}, 
which turns out to be 
a direct wide band-gap semiconductor with potential applications 
in electronic devices and optoelectronics 
\cite{su17b,ch20,dr20,gu18c}. 
Specifically, this material has valuable optical properties as 
large photoluminescence intensity and excitonic effects, due to its 
direct band-gap and electronic quantum confinement \cite{ch20,hs11}.

Several compositions Si$_x$C$_{1-x}$ for 2D silicon carbide have been 
predicted to be also stable, and to behave as semiconductor,
semimetal, or topological insulators, depending on the
stoichiometry \cite{sh15,ch20,fa17}.
The lowest formation energy has been found for the isoatomic
stoichiometry Si$_{0.5}$C$_{0.5}$, which we call 2D SiC \cite{sh15}.

Understanding structural and thermal properties of two-dimensional
systems has been a goal in statistical physics for many years
\cite{sa94,ne04,ta13}, above all in the context of biological
membranes and soft condensed matter \cite{fo08,ta13}.
This problem has expanded its interest to
crystalline membranes, after the synthesis of graphene and
related materials in recent years.
Dealing with crystalline 2D materials allows us to reliably model 
systems at the atomic scale, opening an access to physical 
properties of this kind of systems \cite{po12b,fo13,wa16,he18}.
In this context, electronic structure methods have been used 
since the 1980s to study equilibrium configurations, energetics,
quantum-size effects, and related aspects of ordered
2D systems \cite{mi82,fe84,bo88,bo90,tr92}.

Electronic structure calculations of 2D SiC have shown features
of the minimum-energy configuration for this layered material,
which turns out to be planar.
At finite temperatures, one expects the presence of bending and
ripping in the SiC layer, as has been studied earlier for graphene.
Moreover, quantum effects such as zero-point motion will cause atomic 
delocalization and departure of strict planarity, even at $T = 0$.
Also, nuclear quantum effects can be important for vibrational and 
electronic properties of relatively light atomic species such as 
carbon, as has been shown earlier for graphene, mainly at low 
temperatures.

Path-integral simulations (molecular dynamics and Monte Carlo)
are well suited to appraise effects associated to the quantum
character of atomic nuclei. This kind of simulations allow us
to efficiently quantize the nuclear degrees of freedom,
including both thermal and quantum fluctuations at finite 
temperatures \cite{gi88,ce95}.
This procedure allows one to carry out quantitative analyses
of anharmonic effects in condensed matter \cite{he16,br19}.

In this paper we employ the path-integral molecular dynamics (PIMD)
procedure to study nuclear quantum effects in vibrational,
structural, and thermal properties of 2D SiC in a temperature 
range from from 25 to 1500~K.
Interatomic interactions are obtained from a tight-binding (TB) 
Hamiltonian, built up according to results of calculations based
on density-functional theory (DFT).
Path-integral simulation analogous to those carried out here
were used earlier to analyze nuclear quantum effects in 
carbon-based materials such as diamond \cite{ra06,br20}
and graphite \cite{he21b}, as well as in silicon \cite{no96}.
This kind of techniques have been applied in recent years
to study 2D materials as graphene \cite{br15,ha18,he16}
and BN \cite{br22}.
Here we will discuss similarities and differences of 2D SiC
with other 2D materials, especially graphene.

The paper is organized as follows. In Sec.\,II we explain the
computational procedures used here: PIMD technique and tight-binding
method. In Sec.\,III we give results obtained in a harmonic 
approximation of the vibrational modes in 2D SiC. 
In Sec.\,IV we present results for the internal energy derived
from PIMD simulations, as well as its constituent parts, 
kinetic and potential energy. Interatomic distances and
atomic mean-squares displacements are discussed in Secs.~\,V
and VI, respectively. A discussion on the layer area (in-plane 
and {\em real}) is given in Sec.~\,VII.
The paper closes with a Summary of the main results.

\section{Method of calculation}

\subsection{Path-integral molecular dynamics}

The Feynman path-integral formulation of statistical physics \cite{fe72} 
is a well established tool to study many-body quantum systems at
finite temperatures. This method is nowadays employed to
study properties of condensed matter by means of its implementation
in numerical simulations using Monte Carlo or molecular dynamics
techniques.
In this paper we use the path-integral molecular dynamics (PIMD)
method to study the influence of nuclear quantum effects on
structural and vibrational properties of 2D silicon carbide 
at several temperatures.
In this section we give some details on this computational procedure,
pertinent for the presentation of our results for 2D-SiC. 
More details on this kind of atomistic simulations
are given elsewhere \cite{ce95,gi88,tu10,he14}.

Our simulations are carried out in the isothermal-isobaric ensemble
with variables $N$ (number of particles), $\tau$ (in-plane stress),
and $T$ (temperature). We consider a simulation cell with $N/2$ carbon
and $N/2$ silicon atoms. We take $\tau = 0$, so that the external
stress on the SiC layer vanishes. $\tau$ (with units of
force per unit length), is the conjugate variable
to the in-plane area $A_p$, which in the following will be
the area of the simulation cell on the $(x, y)$ plane.

The partition function $Z(N, \tau, T)$ for the isothermal-isobaric 
ensemble is given by
\begin{equation}
  Z(N, \tau, T) = \int d A_p \, 
	\exp(-\beta \tau A_p )\,Z(N, A_p, T) \, ,
\label{1eq}
\end{equation}
where $Z(N, A_p, T)$ is the canonical partition function, 
$\beta = 1 / (k_B T)$, and $k_B$ is Boltzmann's constant. 
Using the Trotter formula and a high-temperature approximation 
for the density matrix \cite{gi88,kl90,he14}, $Z(N, A_p, T)$ 
can be written for our SiC layer system as
\begin{multline}
  Z(N, A_p, T) \approx
	\left( \frac {M_{\rm C}^{1/2} M_{\rm Si}^{1/2} P}
	{2\pi\beta{\hbar}^2} \right) ^ {3 P N/2}
     \\   \int  d{\bf R}_1 \dots d{\bf R}_P \;
        \exp \left[ -\beta \; V_{\text {eff} }
                ({\bf R}_1,...,{\bf R}_P) \right] \, .
\label{2eq}
\end{multline}
Here, ${\bf R}_p \, (p = 1, ..., P)$ is a $3N$-dimensional vector,
whose components are the Cartesian coordinates of the atomic 
nuclei $({\bf r}_{1,p},...,{\bf r}_{N,p})$.
The index $p$ indicates the path coordinate, which is discretized 
into $P$ points (Trotter number) along a path, and the cyclic 
condition imposes ${\bf R}_{P+1} = {\bf R}_1$.
$M_{\rm C}$ and $M_{\rm Si}$ are the atomic masses of C and Si,
respectively.

Thus, $Z(N, A_p, T)$ is equivalent to the canonical partition 
function of a classical system with an effective potential:
\begin{multline}
V_{\text {eff}}({\bf R}_1,...,{\bf R}_P) =
      \frac {P}{2 \beta^2 {\hbar }^2}
      \sum_{j=1}^{N} \sum_{p=1}^{P}
         M_j ({\bf r}_{j,p+1} - {\bf r}_{j,p})^2  +
   \\ +  \frac1P \sum_{p=1}^P  V({\bf R}_p)  \; ,
\label{3eq}
\end{multline}
where $j$ runs over the $N$ atomic nuclei, $M_j = M_{\rm C}$ for 
$j \leq N/2$, and $M_j = M_{\rm Si}$ for $j > N/2$.
$V_{\text {eff}}({\bf R}_1,...,{\bf R}_P)$ corresponds
to the interaction potential of a classical 
system made up of $N$ cyclic chains (one per atomic nucleus),
where successive elements (beads) are coupled by a harmonic 
interaction with force constant
$\kappa_j = {M_j P}/{\beta^2 {\hbar }^2}$ (first term on the
r.h.s. of Eq.~(\ref{3eq})).
The interchain coupling is restricted to beads with the same 
index $j$, and corresponds to interaction potential 
$V({\bf R}_p)$ for $p = 1, ..., P$.
Here, this potential is derived from a tight-binding (TB)
Hamiltonian, as explained below in Sec.~II.B.
The expression for $Z(N, A_p, T)$ in Eq.~(\ref{2eq}) is exact 
in the limit $P \rightarrow \infty$, and it is valid for
distinguishable particles, which is justified for C and Si nuclei
in the considered 2D crystalline membrane, since the 
overlap of nuclear wave functions is negligible.

To have a roughly constant accuracy for the results
at different temperatures, we have taken a Trotter number which
scales as the inverse temperature. At a given temperature, the 
value of $P$ needed to obtain convergence of the results 
depends on the scale of the vibrational frequencies
in the material.  We have taken $P T = 6000$~K.
For example, a PIMD simulation for $N$ = 112 atoms at $T$ = 50~K
($P$ = 120), requires dealing with $N P$ = 13,440 classical particles.
The finite Trotter number $P$ causes the appearance of a cutoff for 
the vibrational energy. In fact, the largest energy sampled using
$P$ amounts to $E_c \approx \hbar / \epsilon$, with
$\epsilon = \beta \hbar / P$.
This translates into a cutoff for the vibrational frequencies
at $\omega_c = P k_B T / \hbar \approx 4200$ cm$^{-1}$, much larger 
than the frequencies appearing in 2D SiC (see below).

We employed algorithms for performing PIMD simulations 
in the $N \tau T$ ensemble, as those given in the
literature \cite{ma99,tu10}. In particular, we have used staging 
variables to define the bead coordinates, and
the constant-$T$ ensemble has been obtained by coupling chains
of Nos\'e-Hoover thermostats to each staging variable.
In addition, a barostat chain was coupled to the in-plane
area of the simulation cell to keep an in-plane stress 
$\tau = 0$ \cite{tu10,he14}.
The equations of motion were integrated through the reversible
reference system propagator algorithm (RESPA),  allowing us to
use different time steps for the integration of fast and slow
degrees of freedom \cite{ma96}.
On one side, for the atomic dynamics associated to interatomic 
forces, we used a time step $\Delta t$ = 1 fs.
On the other side, we considered a time step $\delta t = \Delta t/4$,
for the evolution of fast dynamical variables, i.e., 
thermostats and harmonic bead interactions.
The equations of motion employed in our simulations, specific for
2D materials, are given in detail elsewhere \cite{ra20}.
The dynamics in this computational method does not correspond
to the real quantum dynamics of the actual particles, but
it is accurate for sampling the true many-body configuration
space, thus giving precise values for time-independent
equilibrium variables of the considered quantum system.

\begin{figure}
\vspace{-0.8cm}
\includegraphics[width=9cm]{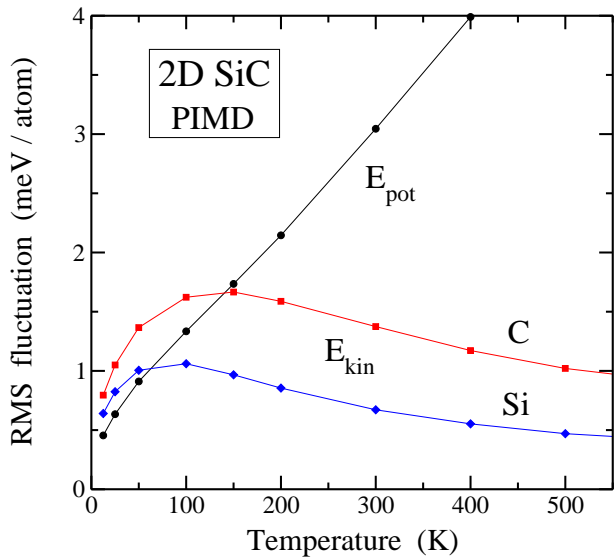}
\vspace{-0.9cm}
\caption{RMS fluctuations of the kinetic and potential energy
for 2D SiC as a function of temperature, as derived from PIMD simulations.
Circles: potential energy; squares: kinetic energy of carbon;
diamonds: kinetic energy of silicon.
Error bars are less than the symbol size.
Curves are guides to the eye.
}
\label{f1}
\end{figure}

The kinetic energy, $E_{\rm kin}$, has been obtained by using the
so-called virial estimator, which has a statistical uncertainty
less than the potential energy of the system, mainly
at high temperature \cite{he82,tu10}. This estimator helps to
determine the mean kinetic energy with good accuracy.
In Fig.~1 we present the root mean-square (RMS) fluctuations 
of the kinetic energy ($E_{\rm kin}$) obtained with this procedure in
our PIMD simulations of 2D SiC, as well as those of the potential
energy ($E_{\rm pot}$) of the system up to 500 K.
On one side,
one observes that RMS fluctuations of $E_{\rm pot}$ grow as $T$ is 
raised, and in fact at high $T$ we find an approximately linear
increase, as expected for thermodynamic fluctuations of a classical
system in the $N \tau T$ ensemble.
On the other side, the RMS fluctuations of $E_{\rm kin}$ obtained
for the virial estimator attain a maximum at about 100~K for Si and 
150~K for C and decrease at higher $T$.

\begin{figure}
\vspace{-4.1cm}
\includegraphics[width=9cm]{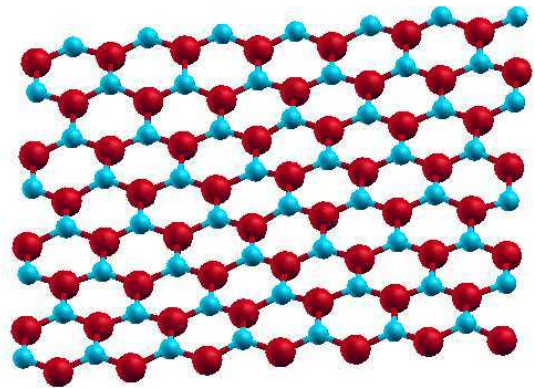}
\vspace{-0.9cm}
\caption{Snapshot taken from a simulation of 2D silicon carbide
at $T = 300$~K. Large red and small light blue spheres indicate
silicon and carbon atoms, respectively.
}
\label{f2}
\end{figure}

For the simulations presented here, we have taken
rectangular simulation cells with similar side length
in the $x$ and $y$ directions of the $(x, y)$ reference plane
($L_x \approx L_y$), where periodic boundary conditions were 
considered. In the out-of-plane $z$-direction, free boundary 
conditions were assumed, so Si and C atoms can move without 
restriction, as in a free-standing layer.
We considered simulation cells with $N$ = 112 atoms,
at temperatures between 25 and 1500~K.
To check the convergence of our results with system size,
some calculations were carried out for $N$ up to 308 atoms.
Given a temperature, a typical simulation run consisted of
$10^5$ PIMD steps for system equilibration, followed by
$4 \times 10^6$ steps for the calculation of ensemble average
properties.  In Fig.~2 we present a view of a 2D SiC configuration
obtained in our simulations at $T = 300$~K.
In this picture, large red and small light blue spheres
represent C and Si atoms, respectively.

To quantify the magnitude of quantum effects in the equilibrium
properties of 2D SiC, some classical molecular dynamics (MD) 
simulations have been also performed with the same TB Hamiltonian.
In our context of path-integral simulations, the classical 
limit is obtained from Eq.~(\ref{3eq}) by putting $P = 1$
(in this case, the first term on the r.h.s. disappears).

\subsection{Tight-binding method}

Our simulations were performed within the adiabatic (Born-Oppenheimer) 
approximation, which allows one to define a potential-energy surface 
for the nuclear dynamics.
We obtain the Born-Oppenheimer surface from an effective tight-binding 
Hamiltonian, based on density functional calculations \cite{po95}.
Thus, our procedure takes into account the quantum nature of both,
electrons and atomic nuclei, the former through the TB Hamiltonian
and the latter by means of path integrals.
In this way, phonon-phonon and electron-phonon interactions are directly 
included in our PIMD simulations.

Total energies and interatomic forces are calculated in our simulations using
the DFT based non-orthogonal TB Hamiltonian of Porezag {\em et al.} \cite{po95}. 
In particular, the TB parametrization for structures containing Si and C  atoms
was presented in Ref. \cite{gu96}. The main steps for this parametrization are:
i) atomic orbitals are derived as the eigenfunctions of appropriately
constructed pseudoatoms, where the charge density of the valence electrons
is concentrated closer to the nucleus; ii) overlap matrices between the
atomic orbitals are  tabulated as a function of the internuclear distance;
iii) matrix elements of the TB Hamiltonian are calculated using DFT in
the local density approximation (LDA), and tabulated also 
as a function of the internuclear distance; iv) finally, a short-range 
repulsive part of the total potential is fitted to self-consistent-field 
LDA data of proper reference systems. The non-orthogonality of the atomic 
basis is an important key for the transferability of the parametrization 
to complex systems \cite{po95}. This TB method is not self-consistent and,
contrary to other empirical TB approaches,
it does not include any temperature effect or any fit to
experimental data in its parametrization.

The TB model used in this paper was employed earlier
to analyze the (1x1) reconstruction of the (110) SiC surface \cite{gu96},
to study nuclear quantum effects in 3C SiC \cite{ra08}, and to investigate 
isotope effects in this 3D material \cite{he09c}. A detailed review on 
the ability of TB methods to precisely describe several properties of 
solids and molecules was presented by Goringe et al. \cite{go97}.

\begin{figure}
\vspace{-0.8cm}
\includegraphics[width=9cm]{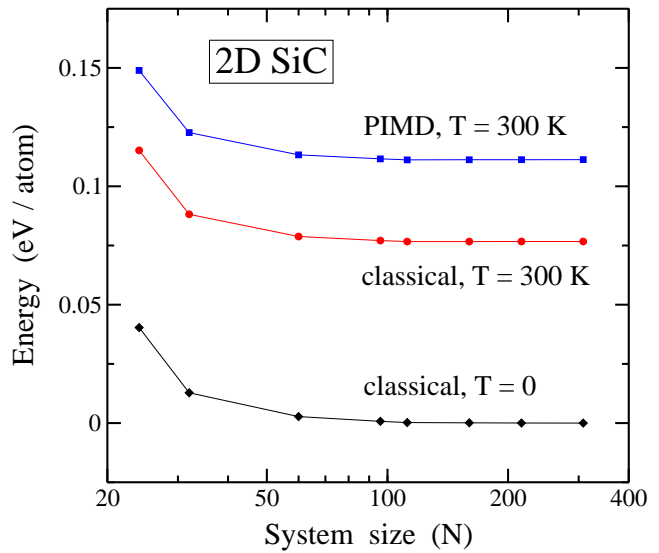}
\vspace{-0.9cm}
\caption{Potential energy of 2D SiC for various simulation cell
sizes ($N$).  Diamonds: classical, $T = 0$;
circles: classical, $T = 300$~K; squares: quantum, $T = 300$~K.
The zero of energy is taken from the $T = 0$ classical result
for $N = 308$. Error bars are less than the symbol size.
Note the logarithmic scale in the horizontal axis.
}
\label{f3}
\end{figure}

For the reciprocal-space sampling of electronic degrees of freedom
we considered only the $\Gamma$
point (${\bf k} = 0$), since the main effect of using larger 
${\bf k}$ sets is a shift in the total energy, with negligible 
effect on the calculation of energy differences.
A similar effect appears for the energy as a function of the
simulation-cell size.
This is shown in Fig.~3, where we display the convergence of 
the potential energy of 2D SiC for several cell sizes $N$. 
The data points correspond to $E_{\rm pot}$ calculated with 
the TB model ($\Gamma$ point) for the minimum-energy configuration 
(classical, $T = 0$).
In this figure, we also present results for the energy 
obtained for several cell sizes from classical MD (circles) and
PIMD simulations (squares) at $T$ = 300~K.
In both cases, we observe a rigid energy shift with respect
to the classical calculations at $T = 0$.

\section{Harmonic approximation}

For the sake of comparison with the results of our PIMD simulations 
of 2D SiC, we present here a harmonic approximation (HA) for 
the atomic vibrational modes. 
Such an approximation in condensed matter is usually rather 
precise at low temperatures. Anharmonicity appears as temperature 
increases, and the outcomes of the HA gradually deviate from
those obtained for more realistic atomistic simulations.
In the HA, vibrational frequencies are assumed to be constant
(independent of the temperature, i.e., those derived for the 
minimum-energy configuration), and changes in the in-plane
area $A_p$ with temperature are not taken into account. 
Thus, we are not considering a quasi-harmonic 
approximation \cite{as76,de96},
where frequencies can change with temperature, which can be
useful for some calculations not addressed here.

\begin{figure}
\vspace{-0.8cm}
\includegraphics[width=9cm]{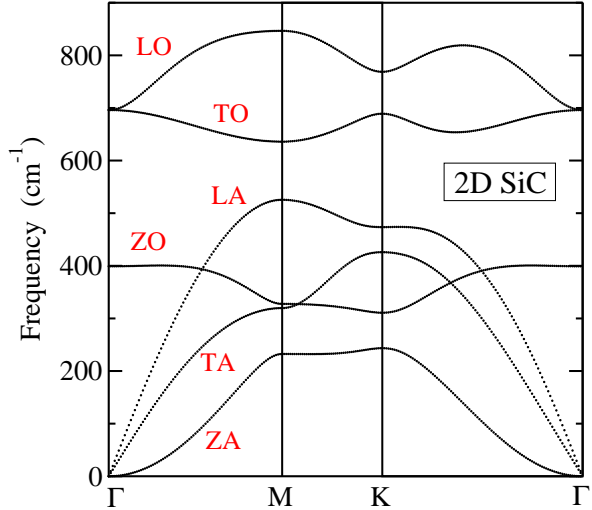}
\vspace{-0.9cm}
\caption{
Phonon dispersion bands of 2D silicon carbide, $\omega_j({\bf k})$,
as derived by diagonalization of the dynamical matrix for the
TB Hamiltonian employed here. Labels indicate the six phonon bands.
}
\label{f4}
\end{figure}

The phonon dispersion of 2D SiC, derived from the TB Hamiltonian
by diagonalization of the dynamical matrix is shown in Fig.~4
along high-symmetry directions of the Brillouin zone.
We obtain six phonon bands, corresponding to two atoms (C and Si)
in the crystallographic unit cell. Labels indicate the usual
names of the phonon bands:
four branches with in-plane atomic displacements (LA, TA, LO, TO,
L = longitudinal, T = transversal, A = acoustic, O = optical),
and two branches with motion along the $z$ direction (ZA and ZO)
\cite{ya08,ko15b,be10b,gu18c}.
It is relevant for our later discussion the presence of 
the flexural ZA band, parabolic close to the $\Gamma$ point, and
typical of 2D materials \cite{ra19}. 

The sound velocities in 2D SiC can be obtained from the derivative
$\partial \omega / \partial k$ for the acoustic phonon bands
close to $\Gamma$ point.
From the LA and TA bands shown in Fig.~4, we find $v_L$ = 13.0 km/s
and $v_T$ = 8.3 km/s.
The elastic stiffness constants of this 2D material, $C_{11}$ and
$C_{12}$, may be derived from the sound velocities and the
surface mass density, $\rho$, as
$C_{11} = \rho v_L^2$ = 9.17 eV/\AA$^2$ and
$C_{12} = C_{11} - 2 \rho v_T^2$ = 1.77 eV/\AA$^2$.
From these constants, one can calculate the 2D modulus of
hydrostatic compression: $B_{xy} = (C_{11} + C_{12}) / 2$
= 5.47 eV/\AA$^2$. This variable is analogous to the bulk
modulus in 3D materials \citep{be96b}, and the value found for 
2D SiC is clearly lower than that corresponding to graphene 
($B_{xy}$ = 12.7 eV/\AA$^2$ \cite{ra19}), as a consequence
of weaker bonds in the former.

The flexural ZA band follows close to $\Gamma$ a quadratic
dependence on $k$:
$\omega_{\rm ZA}(k) = (\kappa / \rho)^{1/2} \, k^2$,
where $\kappa$ is the bending constant.
From the ZA band shown in Fig.~4, we find $\kappa$ = 1.0 eV,
which turns out to be somewhat less than the bending constant
of graphene ($\kappa$ = 1.5 eV) \cite{ra19}.

\begin{figure}
\vspace{-1.3cm}
\includegraphics[width=9cm]{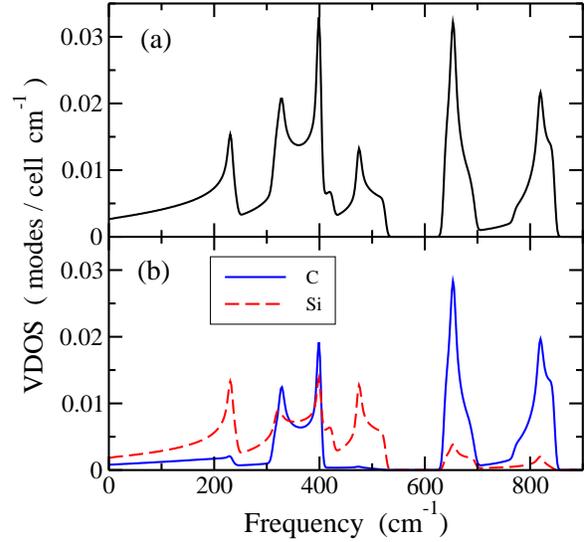}
\vspace{-0.8cm}
\caption{
Vibrational density of states of 2D silicon carbide,
as derived from the HA for the TB Hamiltonian
employed for the PIMD simulations.
(a) Total density of states, $g(\omega)$.
(b) Solid and dashed curves indicate the VDOS
corresponding to carbon, $g_{\rm C}(\omega)$, and
silicon, $g_{\rm Si}(\omega)$, respectively.
}
\label{f5}
\end{figure}

For comparison with results of our PIMD simulations, it is interesting
to obtain the vibrational density of states (VDOS) for the whole 
Brillouin zone. This has been done by numerical integration over 
the hexagonal zone, according to the procedure described in 
Ref.~\cite{ra86}. In Fig.~5(a) we present the resulting VDOS for 
2D SiC, and in Fig.~5(b) we have plotted separately 
the contributions of carbon (solid curve) and silicon
(dashed curve). We will call these contributions $g_{\rm C}(\omega)$
and $g_{\rm Si}(\omega)$, respectively, and their sum yields
the total VDOS, $g(\omega)$.
Note that the density of states converges to a positive value for 
$\omega \to 0$, due to the contribution of flexural ZA modes
(see below).

In a quantum HA, the vibrational energy per atom in a SiC monolayer
is given by
\begin{equation}
 E_{\rm vib}  =  \frac{1}{2N} \sum_{r,\bf k} \hbar \, \omega_r({\bf k})
   \coth \left( \frac12 \beta \hbar \, \omega_r({\bf k}) \right)  \, ,
\label{evib}
\end{equation}
where the index $r$ ($r$ = 1, ..., 6) indicates the phonon bands.
The sum in ${\bf k}$ runs over wavevectors
${\bf k} = (k_x, k_y)$ in the 2D hexagonal Brillouin zone,
with ${\bf k}$ points spaced by $\Delta k_x = 2 \pi / L_x$ and
$\Delta k_y = 2 \pi / L_y$ \cite{he16,ra19}.
In the sequel, $k$ will stand for the wavenumber,
i.e., $k = |{\bf k}|$.

Given a VDOS $g(\omega)$ for the lattice modes, the vibrational 
energy in a continuous approximation is given by
\begin{equation}
   E_{\rm vib}  =  \frac12  \int_0^{\omega_m} \hbar \, \omega
         \coth \left( \frac12 \beta \hbar \, \omega \right)  
	 g(\omega) d \omega  \, ,
\label{evib2}
\end{equation}
where $\omega_m$ is the maximum frequency in the solid.
The normalization condition is
\begin{equation}
  \int_0^{\omega_m}  g(\omega)  d \omega  = 6,
\end{equation}
for the six degrees of freedom in a crystallographic unit cell
(one C and one Si).

To analyze the temperature dependence of the energy at low $T$, 
one can consider the continuous model for frequencies and wavevectors, 
as in the Debye model for vibrations in solids \cite{as76,ki96}.
Calling $E_0$ the minimum energy of the 2D material, the vibrational
energy, $E - E_0$, is controlled at low-temperatures by acoustic modes 
with small $k$, close to the $\Gamma$ point.
In our case of 2D SiC, they are TA and LA modes
with $\omega_r \propto k$, as well as ZA flexural modes with 
$\omega_r \propto k^2$.

In general, for a phonon branch $r$ with dispersion relation
$\omega_r \propto k^n$ for small $k$, the contribution to the energy 
at low $T$ is
\begin{equation}
 E_r - E_r^0  =  \int_0^{k_m} \frac{ \hbar \, \omega_r(k)} 
   {\exp [\beta \, \hbar \, \omega_r(k)] - 1} \, g(k)  \, d k  \, ,
\label{ert}
\end{equation}
where $k_m$ is the maximum wavenumber $k_m = (2 \pi / A_0)^{1/2}$  
and $g(k) = A_0 k / 2 \pi$ for 2D materials.
The dispersion relation $\omega_r(k)$, yields a vibrational 
density of states $\bar{g_r(\omega)} = g(k) d k / d \omega
\sim \omega^{\frac2n - 1}$.
Putting $x = \beta \hbar \, \omega / 2$, one finds
\begin{equation}
 E_r - E_r^0  =  K  
  \left( \frac{k_B T}{\hbar} \right)^{1 + \frac2n}
  \int_0^{x_m}  x^{\frac2n} \, ({\rm e}^x - 1)^{-1}  \, d x  \, ,
\label{cvr3}
\end{equation}
$K$ being a constant.
At low temperatures ($k_B T \ll \hbar \, \omega_m$,
i.e. $x_m \gg 1$), we have $E_r - E_r^0 \sim T^{1+2/n}$,
which means a quadratic dependence of $E_r - E_r^0$ on $T$ for 
the flexural ZA branch ($n = 2$), while $E_r - E_r^0 \sim T^3$ 
for LA and TA bands ($n = 1$).
Considering the constants in the integrals above, we obtain for
the ZA phonon branch
\begin{equation}
 E_{ZA}(T) = E_{ZA}^0 + \frac{\pi A_0}{24 \hbar}
  \left( \frac{\rho}{\kappa} \right)^{\frac12}  k_B^2 \, T^2  \, ,
\label{cvt}
\end{equation}
and for LA and TA modes:
\begin{equation}
  E_{ac}(T) = E_{ac}^0 + 
     \frac{\zeta(3) A_0}{\pi \hbar^2 v^2} \, k_B^3 \, T^3  \, ,
\label{cvt2}
\end{equation}
where $v$ is the sound velocity in each phonon band,
and $\zeta$ is Riemann's zeta function.
This gives for the coefficients of $T^2$ and $T^3$
in the low-temperature expansion of the energy, 
the values $1.3 \times 10^{-7}$ eV/K$^2$ and
$4.4 \times 10^{-10}$ eV/K$^3$, respectively.

\section{Energy}

Here we present and discuss results of the internal energy
of 2D SiC, derived from our simulations in the 
isothermal-isobaric ensemble for vanishing external stress.
At $T = 0$, we find a flat SiC sheet for the minimum-energy 
configuration in a classical calculation with the atoms fixed
at their equilibrium positions,
yielding an energy $E_0 = -41.0534$~eV/atom, which is taken 
as a reference for our calculations at finite temperatures.
For a quantum description of the atomic nuclei, one has
zero-point in-plane and out-of-plane atomic fluctuations,
so that the SiC layer is not totally flat.

\begin{figure}
\vspace{-0.7cm}
\includegraphics[width=7.0cm]{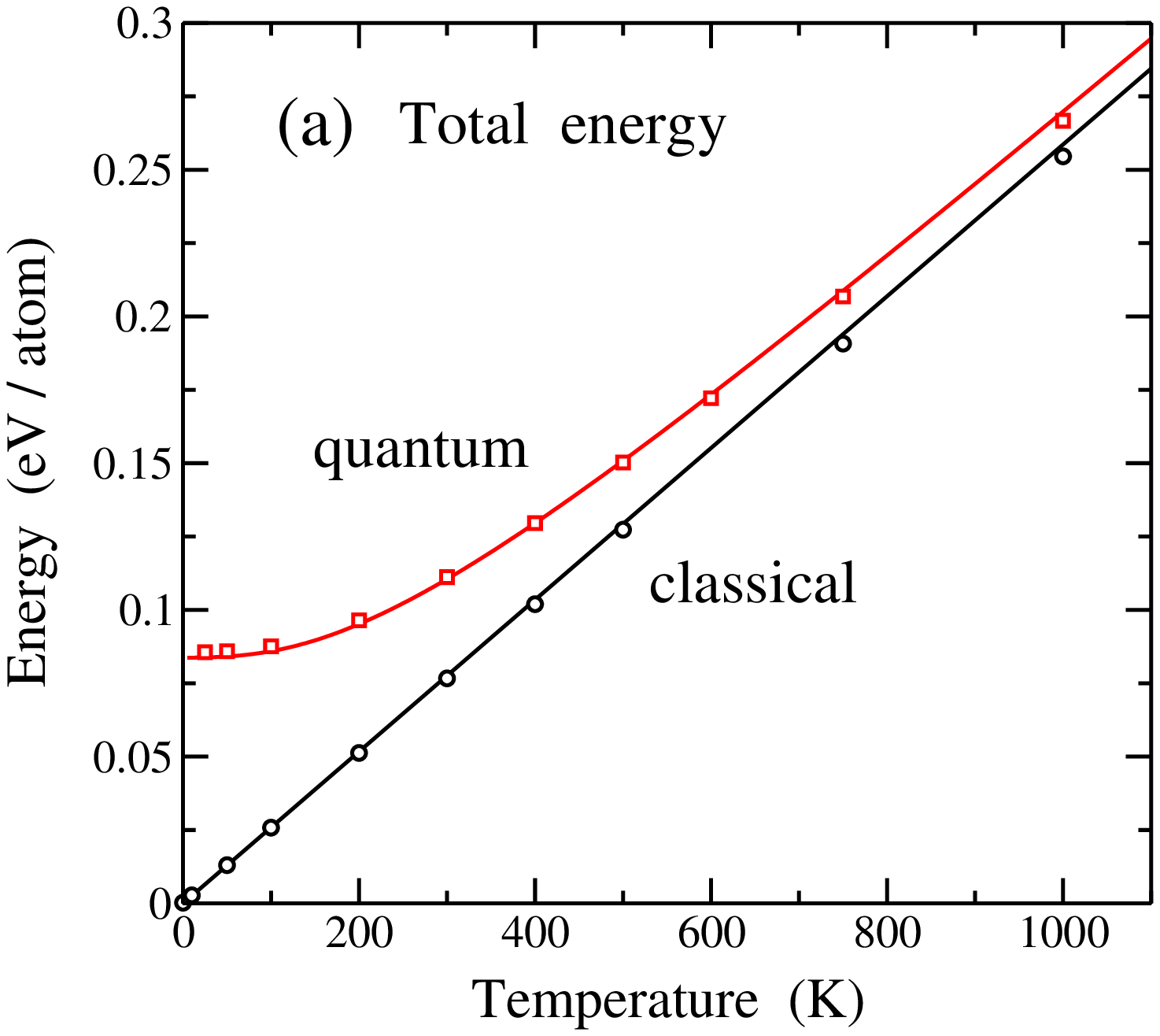}
\includegraphics[width=7.0cm]{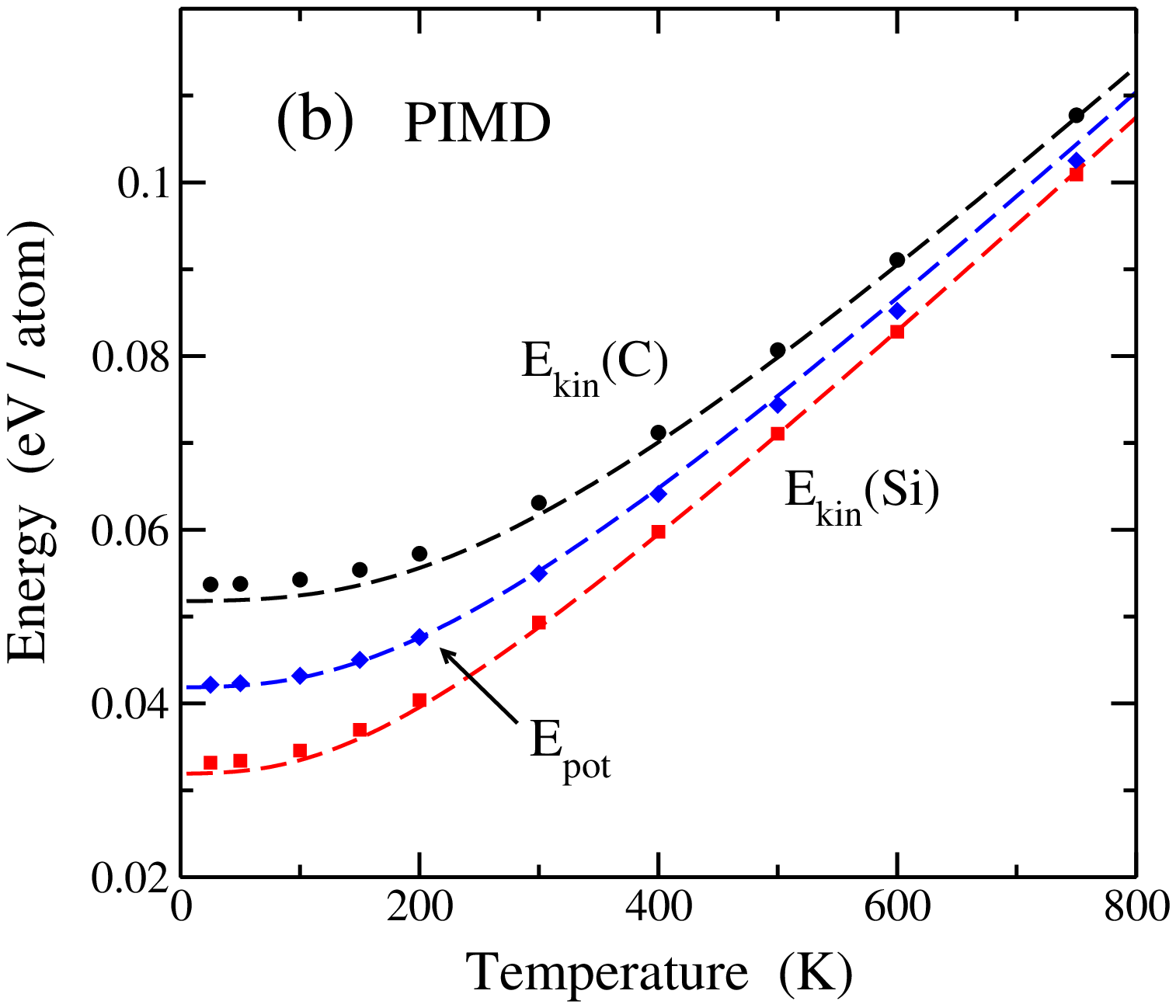}
\vspace{-0.5cm}
\caption{
Temperature dependence of the energy per atom in 2D SiC,
as derived from the HA (curves) and simulations (symbols).
(a) Total energy: circles, classical MD; squares, PIMD.
(b) Different contributions to the total energy:
potential energy (diamonds); kinetic energy of carbon
(circles) and silicon (squares).
Error bars are less than the symbol size.
}
\label{f6}
\end{figure}

In Fig.~6(a) we display the internal energy of 2D SiC as a function
of temperature, obtained from our PIMD simulations (open squares).
For comparison, we also present the internal energy found in 
classical MD simulations (open circles). 
Solid curves indicate the energy obtained in a HA for
both quantum and classical models.
In the quantum case, $E - E_0$ has been calculated by using 
Eq.~(\ref{evib2}) with the VDOS shown in Fig.~5.
The open circles are located near the classical harmonic model, 
i.e., $E^{\rm cl} - E_0 = 3 k_B T$ per atom. In fact, at low $T$, 
they are indistinguishable from the harmonic classical energy.
For $T \gtrsim 400$~K, the simulation results depart progressively
(but slightly) from the harmonic expectancy, and at
temperatures in the order of 1000~K this departure is of
about a 2\%.
For the quantum results, however, the energy derived from PIMD 
simulations is slightly higher than that corresponding to the HA
at low temperature, and for increasing $T$ it
approaches the result of classical simulations. At $T = 1000$~K,
we observe a difference of 12 meV/atom between quantum and
classical data.
For a classical model, at low temperature the atomic motion does
not explore the energy regions far from the absolute minimum,
due to the smallness of the vibrational amplitudes.
For a quantum model, however, the vibrational amplitudes in
the limit $T \to 0$ remain finite, and detect 
anharmonicities in the interatomic potential.

The potential ($E_{\rm pot}$) and kinetic ($E_{\rm kin}$) parts 
of the internal energy $E$ are given separately in PIMD
simulations \cite{he82,tu10,he14}. For our calculations
with external stress $\tau = 0$, one has
$E - E_0 = E_{\rm kin} + E_{\rm pot}$.
In Fig.~6(b), we show the kinetic and potential energy obtained from
PIMD simulations as a function of temperature:
$E_{\rm kin}$(Si) (squares); $E_{\rm kin}$(C) (circles), and 
$E_{\rm pot}$ (diamonds).
Dashed curves correspond to the results of a HA, using Eq.~(\ref{evib2}).
In a harmonic model for the vibrational modes, one has
$E_{\rm kin} = E_{\rm pot}$ (virial theorem \cite{la80,fe72})
for both classical and quantum approaches.
From our results, it is plain that anharmonicity produces an increase
in the kinetic energy of both C and Si at low temperature, while
the potential energy follows closely the harmonic expectation.

For $T \to 0$, we find in the HA $E_{\rm kin}^0$(Si) = 31.9 meV and
$E_{\rm kin}^0$(C) = 51.8 meV. The anharmonic shift amounts to 
1.2 and 1.9 meV/atom for Si and C, which represents an increase of
3.8 and 3.7\%, respectively, as compared to the harmonic calculation.
The slight difference between these relative values is smaller than
the uncertainty derived from the error bars of the kinetic energy
obtained in our simulations.
For the potential energy we obtain $E_{\rm pot}^0$ = 41.8 meV / atom.
We note that $E_{\rm pot}$ cannot be split into C and Si 
contributions, as can be done for the kinetic energy.
The structure of the TB Hamiltonian does not allow identification of
separate contributions to the potential energy for each species.
We observe that $E_{\rm pot}$ derived from PIMD simulations gradually
deviates from the harmonic expectancy as temperature rises.
On the other side, $E_{\rm kin}$ from the simulations approaches the
harmonic result for both C and Si for increasing $T$, as the
low-temperature shift is compensated for by a slower increase
at finite temperatures.

To understand the energy results of PIMD simulations at low $T$,
we note that analysis based on perturbation theory and
quasiharmonic approximations point out that the low-$T$ energy shift
relative to a harmonic model is basically caused by a change of
the kinetic energy \cite{he16,br19}.
This happens, indeed, for perturbed harmonic oscillators at $T = 0$
(considering perturbations of $x^3$ or $x^4$ type), where 
first-order energy changes are due to shifts in $E_{\rm kin}$, 
while $E_{\rm pot}$ stays unmodified respect to the harmonic 
energy \cite{la65}.

\section{Interatomic distances}

In this section we discuss the interatomic distance 
$d_{\rm Si-C}$ between nearest neighbors in 2D SiC.
Zero-point motion is expected to cause an increase in the
equilibrium $d_{\rm Si-C}$. This is a combination of 
quantum dynamics on one side and anharmonicity of the lattice
vibrations on the other. For purely harmonic quantum vibrations,
no change in $d_{\rm Si-C}$ can appear.

\begin{figure}
\vspace{-0.8cm}
\includegraphics[width=9cm]{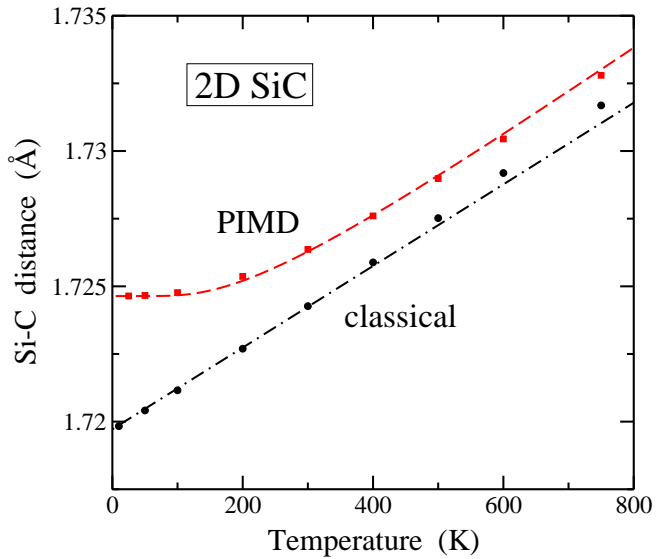}
\vspace{-0.9cm}
\caption{
Temperature dependence of mean interatomic distance,
$d_{\rm Si-C}$ in 2D silicon carbide.
Solid circles and squares represent results of classical and
PIMD simulations, respectively.
Error bars are less than the symbol size.
The dashed curve is a fit of the quantum results to Eq.~(\ref{dd0}).
The dashed-dotted line is a linear fit of the classical
MD results for $T < 500$~K.
}
\label{f7}
\end{figure}

In Fig.~7 we display the mean Si--C distance as 
a function of temperature. Solid squares represent results of
PIMD simulations, whereas circles are data points obtained from
classical MD simulations.
At low temperature ($T \to 0$), we obtain in the quantum simulations
an interatomic distance of 1.725 \AA, 
which increases for rising temperature.
Comparing results for $N$ = 112 and 216 atoms, we find that
size effects of the finite simulation cells on $d_{\rm Si-C}$ are
negligible, as in fact they are smaller than the error bars of
the simulation results (less than the symbol size in Fig.~7). 
For comparison with the results of 2D SiC, we note that PIMD 
simulations of 3C SiC employing the same TB Hamiltonian give a 
zero-temperature distance $d_{\rm Si-C}$ = 1.888~\AA, somewhat 
larger than in the 2D sheet \cite{ra08,he09c}.

At low $T$, the results of classical simulations display a linear
increase with rising temperature, as shown in Fig.~7. This is
typical for interatomic distances in solids from classical
calculations \cite{ki96}.
For $d_{\rm Si-C}$ in 2D SiC we find in the low-temperature 
classical limit a value of 1.720~\AA. 
The dashed-dotted line in Fig.~7 is a linear fit to the
classical results for $T < 500$~K. We observe that the results
of the MD simulations depart progressively from this line at
higher temperatures.

Comparing the low-temperature results for PIMD and classical
MD simulations, we find a zero-point expansion 
$\Lambda_0 = 5 \times 10^{-3}$~\AA\ due to quantum fluctuations,
i.e., an increase in $d_{\rm Si-C}$ of a  0.3\% with respect to 
the classical value.
Note that such an increase in bond length associated to nuclear 
quantum effects turns out to be much larger than the precision 
attained for interatomic distances from diffraction 
experiments \cite{ya94,ra93b,ka98}.
The zero-point bond dilation is similar in magnitude to
the thermal expansion obtained in the classical simulations
from $T= 0$ to 350~K.
The difference between quantum and classical results decreases
as temperature is raised, since they converge to one another
at high $T$, when the relevance of quantum fluctuations decreases.
At $T = 750$~K, this difference is still observable in Fig.~7, 
but it is about 5 times smaller than the zero-point bond expansion.
For the interatomic distance in monolayer graphene, 
a TB Hamiltonian analogous to that employed here 
gives a zero-point bond expansion of a 0.5\%,  
i.e., a relative increase larger than for the larger
Si--C bond in 2D silicon carbide.

Both thermal bond expansion and zero-point dilation
are related to anharmonicity in the interatomic potential,
as in 3D crystalline materials. For 2D SiC, these effects
are mainly associated to anharmonicity of the stretching
vibrations of the Si--C bonds. Something more complex is
needed to describe thermal changes of the in-plane area $A_p$,
because of the coupling between out-of-plane and in-plane
vibrational modes, as discussed below.

A simple quantitative approach to the temperature dependence 
of the Si--C bond length can be obtained in the spirit of 
a quasiharmonic approximation \cite{he21}. 
In this approach one can write
\begin{equation}
   d_Q(T) = d_0 + \frac{\Lambda_0}{E_Q(\omega_{\rm eff},0)}  
	    E_Q(\omega_{\rm eff},T)  \, ,
\label{dd0}
\end{equation}
where $\Lambda_0$ is the zero-point quantum expansion,
$\omega_{\rm eff}$ is an effective frequency, and
$E_Q$ is the harmonic vibrational energy for $\omega_{\rm eff}$. 
In the context of such quasiharmonic approximation, $\Lambda_0$ 
can be interpreted as a ratio $\gamma_{\rm eff} / B_{\rm eff}$ 
between a Gr\"uneisen parameter and an effective compression 
modulus $B_{\rm eff}$ \cite{he21}.
The dashed curve in Fig.~7 shows the interatomic distance
obtained using Eq.~(\ref{dd0}) with an effective frequency
$\omega_{\rm eff}$ = 405~cm$^{-1}$, which follows closely
the results of PIMD simulations in the displayed temperature 
range.

\section{Atomic mean-square displacements}

The PIMD simulations employed here are well-suited to
study atomic delocalization in 3D space at finite temperatures. 
This contains both, a classical (thermal) delocalization,
as well as a contribution due to the quantum character of 
atomic nuclei.
The former is measured by the motion of the center-of-gravity
(centroid) of the ring polymers associated to the quantum particles,
and the latter is given by extension of the quantum paths
(MSD of the beads with respect to their centroid).

\begin{figure}
\vspace{-0.6cm}
\includegraphics[width=7.0cm]{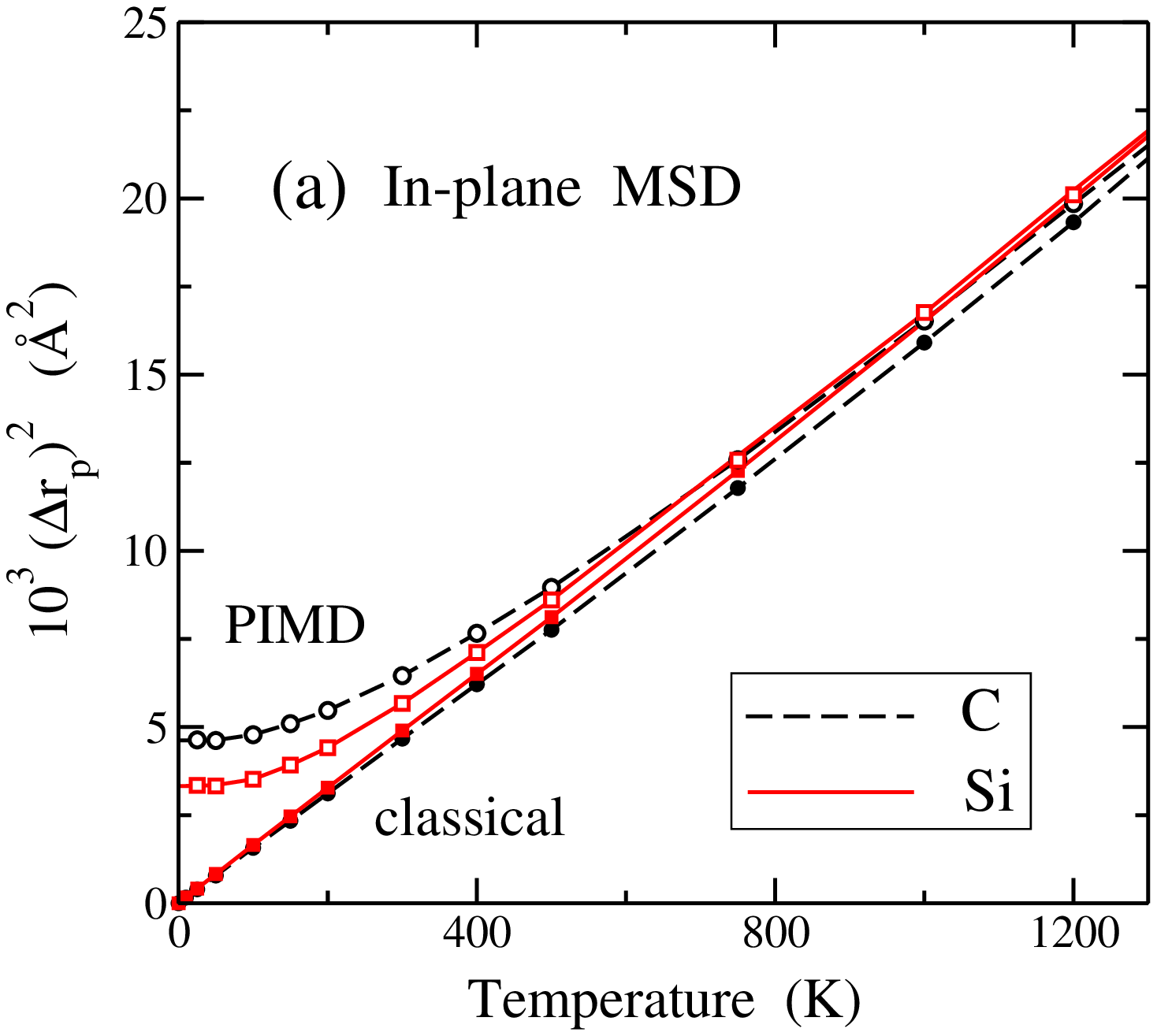}
\includegraphics[width=7.0cm]{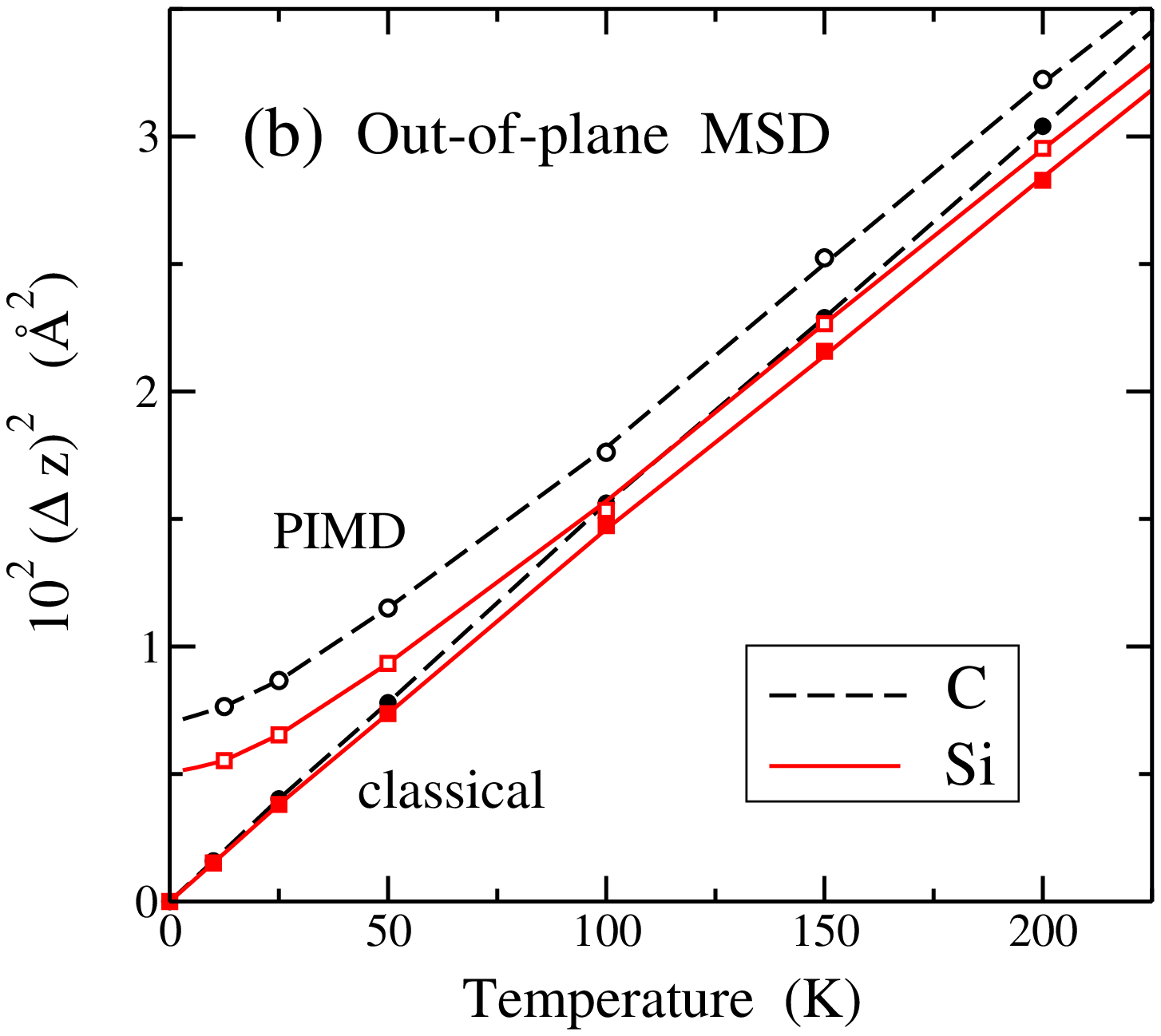}
\vspace{-0.5cm}
\caption{
Atomic mean-square displacements in 2D silicon carbide.
(a) In-plane MSD, $(\Delta r_p)^2$, for carbon (circles)
and silicon (squares) as a function of temperature.
Solid and open symbols indicate results of classical MD
and PIMD simulations, respectively.
(b) Out-of-plane MSD, $(\Delta z)^2$. Symbols and curves
have the same meaning as in (a).
Error bars are in the order of the symbol size.
Curves are guides to the eye.
}
\label{f8}
\end{figure}

We will present separately the in-plane and out-of-plane atomic MSDs. 
In Fig.~8(a) we display the MSD of C and Si atoms on the $(x,y)$ plane.
Symbols are data points obtained from our simulations:
open and solid symbols indicate results of PIMD and classical
MD simulations, respectively.
Circles and dashed curves correspond to carbon, whereas squares
and solid curves represent data for silicon.
We have checked that contributions along the $x$ and $y$ directions 
are indistinguishable within the error bars of our simulation results
(in the order of the symbol size).
The quantum results converge for $T \to 0$ to 
$(\Delta r_p)_0^2 = (\Delta x)_0^2 + (\Delta y)_0^2$ = 
$4.7 \times 10^{-3}$\AA$^2$ and 
$3.4 \times 10^{-3}$ \AA$^2$ for C and Si, respectively.
For the quantum data, we find $(\Delta r_p)^2_{\rm C}$
$> (\Delta r_p)^2_{\rm Si}$ in the whole temperature range displayed
in Fig.~8(a), and the opposite happens for the classical results.

In a classical calculation the atomic MSDs do not depend on the
atomic mass, but they indeed change with the interatomic potential
felt by the atomic nuclei. In our case, the effective potential
felt by Si atoms is somewhat softer than that felt by C atoms,
which causes a larger classical MSD for Si (solid curve) than 
for C atoms (dashed curve).
This effect does also appear in the quantum results, but in this
case it is overwhelmed by the influence of the atomic mass,
i.e., at low temperature the contribution of a mode with 
frequency $\omega$ contribute to the MSD of an atom with mass $M$ 
as $\sim \hbar / 2 M \omega$.
At high $T$ classical and quantum results converge one to the
other for each atomic species. This means that the quantum
MSD in the $(x, y)$ plane for C and Si cross at a certain
temperature higher than 600~K. In fact, this happens at 
$T \approx 800$~K (not shown in Fig.~8(a)).

The atomic motion in the out-of-plane $z$ direction is important
for various properties of 2D materials, since
it is the origin of bending in their sheets.
In Fig.~8(b) we display results for the MSD of C and Si in the
$z$-direction, obtained from PIMD and classical MD simulations.
Symbols and curves have the same meaning as in Fig.~8(a).
Note that values of $(\Delta z)^2$ for C and Si are clearly
larger than the corresponding ones in the layer plane, due to
the contribution of low-frequency flexural modes 
(ZA) close to the $\Gamma$ point.

For the quantum results in the low-$T$ limit, we find
$(\Delta z)_0^2 = 7.2 \times 10^{-3}$~\AA$^2$  and 
$5.1 \times 10^{-3}$~\AA$^2$, for C and Si respectively.
The ratio between these values is close to the inverse
square root of the mass ratio, i.e., 
$(\Delta z)_{\rm C}^2 / (\Delta z)_{\rm Si}^2$
$\approx  (M_{\rm Si} / M_{\rm C})^{1/2}$.
For the out-of-plane motion, the classical MSD for C is
somewhat higher than that of Si, contrary to the findings for 
in-plane MSDs shown above. This is related to details of
the effective potential felt by each atomic species,
which shows different behavior in out-of-plane and 
in-plane directions.

From earlier simulations of graphene and other 2D materials 
\cite{he16} it is known that, although atomic MSDs in the layer 
plane are rather insensitive to the system size, out-of-plane 
MSD have a size effect, in particular at high temperatures. 
This is due to the presence of long-wavelength vibrational modes
with low frequency and large vibrational amplitudes in the
ZA flexural band.  This phonon branch may be described at finite 
temperatures by a dispersion relation of the form
$\rho \, \omega({\bf k})^2 = \sigma k^2 + \kappa k^4$, where
$\sigma$ is an effective stress \cite{he16,ra19}.
For our present purposes, $\sigma$ is negligible and the
ZA branch can be considered as parabolic:
$\omega({\bf k}) \approx \sqrt{\kappa/\rho} \, k^2$
with $\kappa$ = 1.0 eV.
Vibrational modes with longer wavelength $\lambda$ appear
for increasing system size $N$.  This means that one has
an effective cut-off $\lambda_{max} \approx L$, where
$L = (N A_p)^{1/2}$.
Then, we have $k_{min} = 2 \pi / \lambda_{max}$, so that
$k_{min} \sim N^{-1/2}$, which yields a frequency
$\omega_{min} \sim N^{-1}$.
From the contributions of the whole bands, $(\Delta z)^2$ 
for a given atomic species scales as $N^{\eta}$, with 
an exponent $\eta > 0$.
A precise estimation of $\eta$ for 2D SiC would require carrying 
out longer simulations (millions of simulation steps) 
with cell sizes much larger than those considered here. 
Our present procedure with a TB Hamiltonian does not allow us 
to perform such simulations for $N \gtrsim 400$~atoms in
moderately long computing times.
We also note that the relative statistical uncertainty (error bar)
derived from PIMD simulations depends on the physical variable
at hand, and it is in particular relatively large for the
in-plane area $A_p$.

The consistency of the results of our quantum simulations
in the coordinate and momentum space can be checked from 
the atomic MSDs and kinetic energy given above.
In general, for a particle described by quantum mechanics,
the RMS displacements of the coordinate $x$ and momentum $p_x$ 
have to comply with the Heisenberg's uncertainty relation 
$\Delta x \, \Delta p_x \geq \hbar/2$ (see, e.g., Ref.~\cite{co77}),
so that
\begin{equation}
   (\Delta p_x)^2  \geq  \frac{\hbar^2}{4 (\Delta x)^2}  \, .
\label{dpx2}
\end{equation}
and similar expressions apply for $y$ and $z$ coordinates.
For the atomic nuclei considered here, we have
$\langle p_x \rangle = 0$, so that
$(\Delta p_x)^2 = \langle p_x^2 \rangle$, and the kinetic energy
of a particle with mass $M$ can be written as
\begin{equation}
 E_{\rm kin} = \frac{\langle {\bf p}^2 \rangle}{2 M} =  \frac{1}{2 M}  
   \left[ (\Delta p_x)^2 + (\Delta p_y)^2 + (\Delta p_z)^2 \right] \, .
\label{ekin}
\end{equation}
Then, using Eqs.~(\ref{dpx2}) and (\ref{ekin}), we find
\begin{equation}
  E_{\rm kin}  \geq  F  \equiv
      \frac{\hbar^2}{8 M} \left[ 
      (\Delta x)^{-2} + (\Delta y)^{-2} + (\Delta z)^{-2} \right] \, .
\label{ekinf}
\end{equation}
where $F$ is a function of the atomic MSDs.
Thus, one has a lower bound for the kinetic energy of the particle,
defined from its delocalization in real space.

\begin{figure}
\vspace{-0.8cm}
\includegraphics[width=9cm]{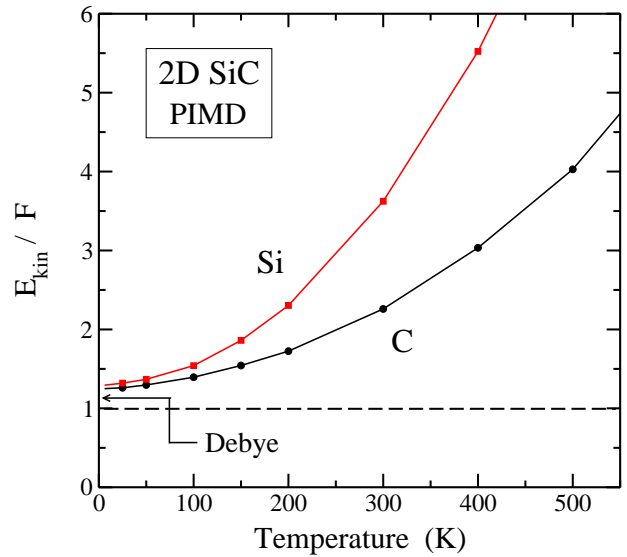}
\vspace{-0.9cm}
\caption{
Temperature dependence of the ratio $E_{\rm kin} / F$ between the
kinetic energy and the function $F$ defined in Eq.~(\ref{ekinf}).
Symbols represents results derived from PIMD simulations
for carbon (circles) and silicon (squares).
Solid curves are guides to the eye.
An arrow indicates the value expected in a Debye approximation
for the vibrational modes.
}
\label{f9}
\end{figure}

In Fig.~9 we present the ratio $E_{\rm kin} / F$ as a function
of temperature for carbon (circles) and silicon (squares), 
as derived from our PIMD simulations of 2D SiC.
The dashed line indicates the lower bound allowed by the uncertainty
relations, i.e., $E_{\rm kin} / F = 1$.
For the sake of comparison, we note that for an isotropic 3D 
harmonic oscillator with frequency $\omega$,
the MSD in each direction ($x$, $y$, and $z$) for the ground state
is $\hbar / 2 M \omega$, and the kinetic energy 
$(E_{\rm kin})_0 = 3 \hbar \, \omega / 4$  \cite{co77}.
This means that the ratio $E_{\rm kin} / F$ converges to unity 
in the limit $T \to 0$.
For atomic motion in condensed matter, one has a frequency 
dispersion, which can be represented by an isotropic 
3D Debye model \cite{ki96,as76}, with a vibrational density of 
states $\mu(\omega) \propto \omega^2$ and a high-frequency
cutoff $\omega_D$. In this case, considering harmonic vibrations,
one finds for $T \to 0$ a ratio $E_{\rm kin} / F = 1.125$,
somewhat higher than for a single harmonic oscillator \cite{he20b}.
This value for the Debye model is indicated in Fig.~9 by an arrow,
a little below the results of our simulation for carbon and silicon
in anisotropic 2D SiC.

\section{In-plane vs real area}

In our simulations in the isothermal-isobaric ensemble, 
one fixes $N$, $T$, and the applied 2D stress in the $(x, y)$ plane
(here, $\tau = 0$), permitting changes in the area $A_p$ of 
the simulation cell. 
This area is a practical variable to perform atomistic simulations 
of 2D materials, and has been studied before in various 
works as a function of temperature and external 
stress \cite{ga14,br15,za09,lo16}. It is not, however, a variable 
to which one can attribute properties of a real material surface,
but a projected area on the reference $(x, y)$ plane.
In our case,
Si and C atoms can move out-of-plane in the $z$ direction, and
measuring the {\em real} surface of the SiC layer will give 
values larger than the in-plane area of the 2D simulation cell.

Then, it is interesting to deal with an additional surface defined 
from the atomic positions along a simulation run.
We consider a {\em real} surface $A$
in 3D space for 2D SiC, obtained from the actual geometry of 
the layer \cite{he18}.  This area $A$ is calculated 
by a triangulation defined from the actual atomic positions
along the simulations. The contribution of each structural
hexagon is obtained as a sum of six triangular areas.
Each triangle is built up from the coordinates of neighboring
Si and C atoms and the center (mean position) of 
the hexagon \cite{ra17}.
One can use other definitions for a real area, as those based
on interatomic distances, which yield results similar to that
considered here \cite{ha16,he16}.

\begin{figure}
\vspace{-0.8cm}
\includegraphics[width=9cm]{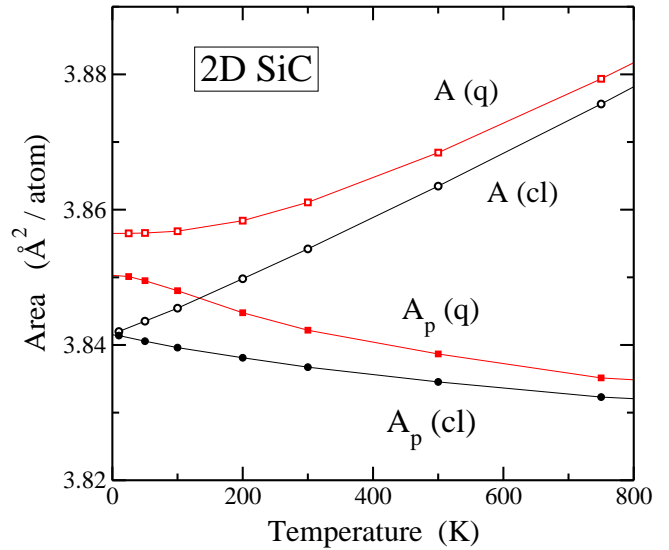}
\vspace{-0.9cm}
\caption{
In-plane area, $A_p$ (solid symbols), and real area per atom,
$A$ (open symbols), vs the temperature.
Circles and squares correspond to classical MD and PIMD simulations,
respectively.
The labels ``q'' and ``cl'' refer to quantum and classical areas.
Error bars are smaller than the symbol size.
Curves are guides to the eye.
}
\label{f10}
\end{figure}

In Fig.~10 we present the temperature dependence of the areas 
$A$ and $A_p$.
In both cases, we show results from PIMD (squares) and classical 
simulations (circles). Solid and open symbols correspond to
the areas $A$ and $A_p$, respectively.
In the classical limit, the surfaces $A_p$ and $A$ take the
same value for $T \to 0$, since the real surface becomes planar
with vanishing out-of-plane atomic displacements.
This area corresponds to the minimum-energy configuration,
$A_0 = 3.8415$~\AA$^2$/atom.
In the quantum results, however, $A_p$ and $A$ do not converge
to the same low-temperature limit. In fact, for $T \to 0$, $A$ 
is larger than $A_p$ by $6 \times 10^{-3}$ \AA$^2$/atom.
Such a difference in the low-$T$ limit appears because the
SiC sheet is not totally planar, because of the zero-point
motion in the $z$ direction.

Apart from differences in the low-temperature behavior, 
a clear feature distinguishing $A_p$ and $A$ is their 
behavior as a function of temperature.
One observes first that the surface $A$ is larger than $A_p$, 
and the difference between both rises as temperature increases.
This is consistent with the fact that
$A_p$ is a projection of $A$ on the $(x, y)$ plane, and 
the real surface becomes progressively bent as temperature rises
and atomic motion in the $z$ direction becomes more relevant.
The in-plane area $A_p$ is found to decrease for increasing $T$
($d A_p / d T < 0$) in both classical and quantum simulations.
$A_p$ reaches a minimum for $T \approx 1400$~K (not shown in Fig~10), 
and slowly increases at higher temperature, 
similarly to the results found for graphene \cite{he16}.
For the real area $A$, however, we find $d A / d T > 0$
in the whole temperature range considered here, for both classical
and PIMD simulations.
Note that the temperature derivative of $A$ and $A_p$ converges
to zero in the low-$T$ limit, as required by the third law
of thermodynamics.

To explain the temperature dependence of the in-plane area 
$A_p$, we observe that it is governed by two main factors.
First, the real area rises for increasing $T$, which is associated
with an increase of its projection on the $(x, y)$ plane, i.e.
the area $A_p$.  Second, bending or rippling of the SiC layer
gives rise to a lowering of the in-plane area. 
The second factor dominates in the temperature region
shown in Fig.~10, causing a decrease in $A_p$ for temperatures 
lower than 1400 K. 
At high temperatures, the first factor (expansion of $A$) 
dominates and one has $d A_p / d T > 0$.
This behavior of $A_p$ for 2D SiC is similar to that described
for graphene \cite{ga14,mi15b,he16}, but for the latter the
difference between classical and quantum results is about 
two times larger than for silicon carbide.

The difference between in-plane and real area has been denoted
{\em hidden} area for graphene \cite{ni17}
and {\em excess} area for fluid membranes \cite{he84,fo08}.
We define, for each temperature $T$, the dimensionless
{\em excess} area of a crystalline membrane 
as $\Omega = (A - A_p)/A_p$.
In an analytical formulation of membranes in the continuum limit,
the relation between $A$ and $A_p$ can be written 
as \cite{im06,wa09,ra17}
\begin{equation}
  A = \int_{A_p} dx \, dy \, \sqrt{1 + (\nabla h(x,y))^2}  \; ,
\label{aap}
\end{equation}
where $h(x,y)$ is the height of the surface, or the distance to
the mean $(x, y)$ plane of the sheet.
The difference $A - A_p$ may be calculated by expanding
$h(x, y)$ as a Fourier series with wavevectors ${\bf k}$
in the 2D Brillouin zone \cite{sa94,ch15,ra17}.
One obtains
\begin{equation}
   A =  A_p  \left[ 1 + \frac{1}{2 N}  \sum_{\bf k}  k^2  
             \langle |H({\bf k})|^2 \rangle  \right]   \; ,
\label{aa}
\end{equation}
where $H({\bf k})$ are the Fourier components of $h(x, y)$.
Thus, the excess area can be written as
\begin{equation}
 \Omega =  \frac{1}{2 N}
    \sum_{r,{\bf k}}  k^2  \langle |\xi_r({\bf k})|^2 \rangle  \; ,
\label{omega}
\end{equation}
where the sum in $r$ is extended to vibrational modes with
$z$ polarization, i.e., ZA and ZO, and $\xi_r({\bf k})$ are the
vibrational amplitudes in branch $r$.
The contribution of ZO modes is negligible vs. that of 
low-frequency ZA modes for ${\bf k}$ close
to the $\Gamma$ point (small $k$).

In a harmonic approximation, the contribution of C atoms to 
$\langle |\xi_{\rm ZA}({\bf k})|^2 \rangle$ is given by
\begin{equation}
 \langle |\xi_{\rm ZA}^{\rm C}({\bf k})|^2 \rangle  =
  \frac{\hbar}{2 M_{\rm C} \, \omega_{\rm ZA}({\bf k})}
  \coth \left[ \frac12 \beta \hbar \, \omega_{\rm ZA}({\bf k}) 
	\right] \; ,
\label{xik}
\end{equation}
and similarly for the Si contribution.
Putting $\omega_{\rm ZA} = \sqrt{\kappa/\rho} \, k^2$, and
using the continuous approximation for wavenumbers as in Sec.~III
for the vibrational energy, we have for the low-$T$ limit of 
the excess area:
\begin{equation}
 \Omega_0 = \frac{\hbar}{4} \left( \frac{\rho}{\kappa} \right)^{\frac12}
 \int_0^{\omega_m}
 \left[ \frac{g_{\rm ZA}^{\rm C}(\omega)}{M_{\rm C}} +
 \frac{g_{\rm ZA}^{\rm Si}(\omega)}{M_{\rm Si}}  \right] d \omega  \, .
\label{omega2}
\end{equation}
Introducing into this expression the VDOS corresponding to the ZA branch
for C and Si, we find a value $\Omega_0 = 1.6 \times 10^{-3}$, which
coincides with that obtained for the areas $A$ and $A_p$
from PIMD at low temperature (see Fig.~10).
At high temperature, the excess area can be approximated
by the classical limit in Eq.~(\ref{omega}), where the
contribution of atoms with mass $M$ to
$\langle |\xi_{\rm ZA}({\bf k})|^2 \rangle$ is given
by $k_B T / M [\omega_{\rm ZA}({\bf k})]^2$.
This yields a linear increase in $\Omega$ as a function of temperature 
for high $T$.  In this classical limit, $\Omega$ vanishes for $T \to 0$.

\begin{figure}
\vspace{-0.8cm}
\includegraphics[width=9cm]{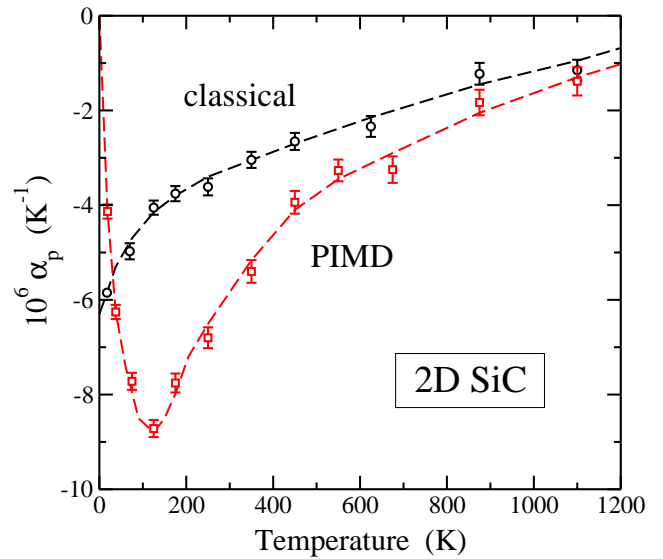}
\vspace{-0.9cm}
\caption{In-plane thermal expansion coefficient $\alpha_p$
vs temperature, as derived from classical (circles) and PIMD
simulations (squares) of 2D silicon carbide.
Symbols are data points obtained from numerical
derivatives of the in-plane area $A_p$.
Curves are guides to the eye.
}
\label{f11}
\end{figure}

In connection with the area $A_p$, one can define an in-plane thermal
expansion coefficient (TEC) as
\begin{equation}
 \alpha_p = \frac{1}{A_p}
      \left( \frac{\partial A_p}{\partial T} \right)_{\tau}   \, .
\end{equation}
This TEC has been studied earlier for 2D materials, and in particular
for graphene \cite{he18,ji09,yo11,ba09}.
In Fig.~11 we present $\alpha_p$ calculated from our simulation
results for 2D SiC up to $T$ = 1200~K.
Open squares are data points found from a numerical derivative
of the area $A_p$ obtained in PIMD simulations.
For comparison we also show results for $\alpha_p$ derived
from classical MD simulations (open circles).
The quantum results display a minimum of $-8.7 \times 10^{-6}$~K$^{-1}$ 
at a temperature $T_m \approx 130$~K. At higher $T$, $\alpha_p$
decreases in absolute value and eventually reaches zero for 
$T \sim$ 1400~K.
At low $T$, $\alpha_p$ approaches zero, as required by
the third law of thermodynamics.
The value of the minimum $\alpha_p$ for SiC is close to the result
reported earlier for graphene: $-9.2 \times 10^{-6}$~K$^{-1}$,
whereas $T_m$ for the former is somewhat lower than for the latter
($\approx 200$~K).

We note that the results for $\alpha_p$ derived from classical
simulations approach to the quantum data at high $T$, but clearly
depart from the PIMD results at low temperature. In fact, the classical
TEC converges to $-6 \times 10^{-6}$ K$^{-1}$ for $T \to 0$,
and does not vanish in this limit, which is a well-known limitation 
of classical models at low temperature.

One can also define a TEC $\alpha = (\partial A/ \partial T) / A$
for the real area $A$ of 2D SiC.
The real area behaves as a function of $T$ in a similar fashion
to the crystal volume of most 3D solids \cite{as76},
i.e., it increases at all finite temperatures.
Then, the TEC $\alpha$ is positive for all temperatures, as 
shown earlier for graphene \cite{he18}.

\section{Summary}

In this paper, we have presented and discussed results of PIMD 
simulations of 2D silicon carbide in the isothermal-isobaric 
ensemble ($N \tau T$) in a temperature range from 25 to 1500~K.
The dynamics of this layered material displays typical 
characteristic of membranes, and an atomic-scale analysis has given 
us insight into the relation between its vibrational modes 
with structural and thermal properties at finite temperatures.
The main focus of our research has been an assessment of
nuclear quantum effects, which can be made by a comparison
of the quantum results with those obtained from classical MD 
simulations.

The use of a TB Hamiltonian to describe the interatomic 
interactions along with PIMD simulations to take into
account the nuclear quantum delocalization has allowed us 
to study 2D silicon carbide, with relatively light 
constituent atoms. 
Our findings show that explicitly taking into account 
the quantum character of atomic nuclei gives appreciable
corrections to the classical results, especially at low 
temperatures, where classical simulations fail to yield the
correct behavior of physical observables, required by the 
third law of thermodynamics.
For $T$ near room temperature, nuclear quantum effects
are still nonnegligible.

The interatomic distance $d_{\rm Si-C}$ and the in-plane area
expand with respect to the classical expectation, which is
a joint signature of zero-point motion and anharmonicity 
of the interatomic potential.  For $T \to 0$, 
the mean Si--C bond and $A_p$ grow by about a 0.3\%.
The thermal expansion of the in-plane area derived from PIMD 
simulations turns out to be negative for $T \lesssim$ 1400~K, 
and positive for higher $T$.  The thermal contraction of 
$A_p$, i.e.  $\alpha_p < 0$, is caused by an increasing 
amplitude of out-of-plane vibrations (mainly ZA modes) as 
temperature is raised.
The real area $A$, however, has a positive thermal expansion,
$\partial A / \partial T > 0$, in the whole temperature range 
considered here, and the difference $A - A_p$ grows for 
rising $T$.

We have quantified the anharmonicity of the vibrational modes
by comparing results derived from a pure harmonic approximation
(VDOS in Fig.~5) with those given by PIMD simulations.
An additional assessment of anharmonicity is obtained from
the difference between the total kinetic and potential energy of 
the system found in the simulations, since they should coincide 
for harmonic vibrations.
At low $T$, we obtain for the kinetic energy of both C and Si
an anharmonic shift of 4\% with respect to the HA , whereas
the potential energy of the layer is not affected by anharmonicity
(as in first-order perturbation theory).

We have found for the bending constant $\kappa$ of 2D SiC a value 
of 1.0 eV, smaller than that corresponding to graphene 
($\kappa$ = 1.5 eV), indicating a larger flexibility of the
former to bend and ripple at finite temperatures.
For the 2D modulus of hydrostatic compression we have found 
$B_{xy}$ = 5.5 eV/\AA$^2$ for SiC vs 12.7 eV/\AA$^2$ for 
graphene, indicating a larger rigidity of the latter in
the layer plane. It will be interesting to research how these
constants evolve from one material to the other, by studying
2D Si$_x$C$_{1-x}$ in a wide composition range, as well as
to consider their changes due to nuclear quantum motion.  

Finally, we note that PIMD simulations in the 
isothermal-isobaric ensemble for tensile and compressive 
stress ($\tau \neq 0$) can give additional information on 
structural and mechanical properties of 2D SiC layers far 
from the minimum-energy configuration.
This kind of simulations may provide insight into the stability 
of this material in a stress-temperature phase diagram.  \\ \\

\noindent
{\bf Data availability} \\

The data that support the findings of this study are available
from the corresponding author upon reasonable request.  \\ \\

\noindent
{\bf Author contribution statement} \\ 

Carlos P. Herrero: Data curation, Investigation, Validation, Original draft

Rafael Ram\'irez: Methodology, Software, Investigation, Validation  \\ \\

\noindent
{\bf Declaration of Competing Interest} \\ 

The authors declare that they have no known competing financial
interests or personal relationships that could have appeared to
influence the work reported in this paper.  \\

\begin{acknowledgments}
This work was supported by Ministerio de Ciencia e Innovaci\'on
(Spain) through Grant PGC2018-096955-B-C44.
\end{acknowledgments}



\end{document}